\documentclass[
aip,%
pof,%
twocolumn,%
reprint,%
floatfix
]{revtex4-1}

\usepackage[english]{babel}
\usepackage[utf8]{inputenc}
\usepackage[caption=false]{subfig}
\usepackage{amsmath}
\usepackage{graphicx}
\usepackage[space-before-unit,range-units = repeat]{siunitx}

\graphicspath{{graphics/}}

\DeclareRobustCommand{\txt}[1]{\text{#1}}
\DeclareRobustCommand{\brc}[1]{\left( #1 \right)}

\begin{document}

\title{Simulations of splashing high and low viscosity droplets}

\author{Arnout M.P. Boelens}
\email{boelens@stanford.edu}
\affiliation{Department of Energy Resources Engineering, Stanford University,
367 Panama Street, Stanford, California 94305, USA}

\author{Juan J. de Pablo}
\affiliation{Institute for Molecular Engineering, The University of Chicago, 5801
South Ellis Avenue, Chicago, Illinois 60637, USA}


\date{\today}

\begin{abstract}
In this work simulations are presented of low viscosity ethanol and high
viscosity silicone oil droplets impacting on a dry solid surface at atmospheric
and reduced ambient pressure. The simulations are able to capture both
the effect of the ambient gas pressure and liquid viscosity on droplet impact
and breakup. The results suggests that the early time droplet impact and gas film
behavior for both low and high viscosity liquids share the same physics.
However, for later time liquid sheet formation and breakup high and low
viscosity liquids behave differently. These results explain
why for both kinds of liquids the pressure effect can be observed, while at the
same time different high and low viscosity splashing regimes have been
identified experimentally.
\end{abstract}

\maketitle

\section{Introduction}

A high and a low viscosity splash on a dry solid surface look rather different. For low viscosity liquids a corona splash is
commonly observed \cite{yarin2006}, while high viscosity liquids initially seem
to spread smoothly over the surface before eventually also breaking up
\cite{stevens2014b}. However, despite these apparent differences, when the ambient gas pressure is reduced splashing is
suppressed for both high and low viscosity liquids \cite{xu2005}. This suggests that there is common physics
behind both types of splashes. This work explores what the difference between
high and low viscosity splashes tells us about splashing in general, and more
specifically about the effect of the ambient pressure on splashing.

The question of whether high and low viscosity splashes
share the same physics is directly relevant to experimental work on
splashing. Because of the very short time scales at which low viscosity
splashes occur, from an experimental point of view it is beneficial to work with
high viscosity liquids and have the physics of splashing play out in slow
motion. For this approach to be valid it is important to know whether results
from low and high viscosity splashes can be compared directly.
For computer simulations of splashing the reverse case can be made.
Due to the computational cost associated with the long time scales of a high
viscosity splash it is much more convenient to simulate a low viscosity splash.
The simulation of a low viscosity ethanol splash takes on the order of several
weeks while a simulation of a high viscosity silicone oil splash takes several
months.

Apart from viscosity there are many more parameters that affect the outcome of a
droplet impacting on a solid surface. These include impact velocity, droplet
diameter, surface tension, and surface roughness \cite{yarin2006,latka2012}.
Despite this wide range of parameters, the effect of a
reduction of the ambient gas pressure is the same: splashing is suppressed
\cite{stevens2014b}. There are various theories that aim to describe droplet
impact and the pressure effect. \citet{mongruel2009} propose that the
characteristic time and length scale of the onset of splashing is determined by a
balance between inertial and viscous forces. On
the other hand, \citet{mandre2009} derive that in the case of an inviscid liquid a droplet initially
skates on a very thin gas film and that the onset of splashing depends on a
combination of the viscosity of this gas film and the parameters of the droplet
impact. According to \citet{riboux2014,riboux2015,riboux2017} whether a droplet splashes
depends on the balance between a lift force acting on the liquid and the growth
rate of the rim at the edge of the droplet. \citet{liu2015} attribute
splashing to a Kelvin-Helmholtz instability of the gas film under the droplet,
and lastly, \citet{sprittles2017} proposes that in the subcontinuum conditions
present in the very thin gas film under the droplet a contact line moves faster
at reduced ambient gas pressure, causing the gas film to close/collapse earlier
at low pressure.

While, none of the above models considers a different high and low viscosity
splashing regime, there is experimental evidence that two different regimes
exist \cite{xu2007b,driscoll2010,stevens2014b}. When plotting the ambient threshold pressure, above which splashing can be
observed, as function of the viscosity of the liquid two distinct regimes can be
observed; for low viscosities the threshold pressure decreases with increasing
viscosity while this trend is reversed for higher viscosities
\cite{xu2007b,driscoll2010,stevens2014b}. In this work computer simulations are presented for ethanol
and silicone oil droplets impacting a dry solid surface at both atmospheric and
reduced ambient pressure. Comparing these two high and low viscosity liquids it
is found that the initial deposition stage for high and low viscosity liquids is
the same, which explains why for both kinds of liquids the pressure effect can
be observed. However, at later times the mechanism for the breakup of the
droplet is significantly different. This is consistent with the experimental
observation of two different splashing regimes.

\begin{figure*}[htbp]
\centering
\subfloat{\includegraphics[width=0.49\textwidth]{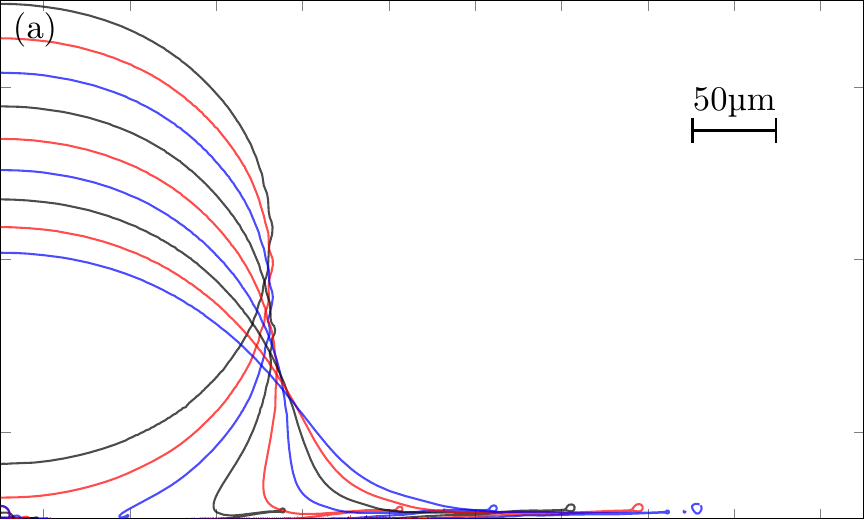}}
\hspace{4pt}
\subfloat{\includegraphics[width=0.49\textwidth]{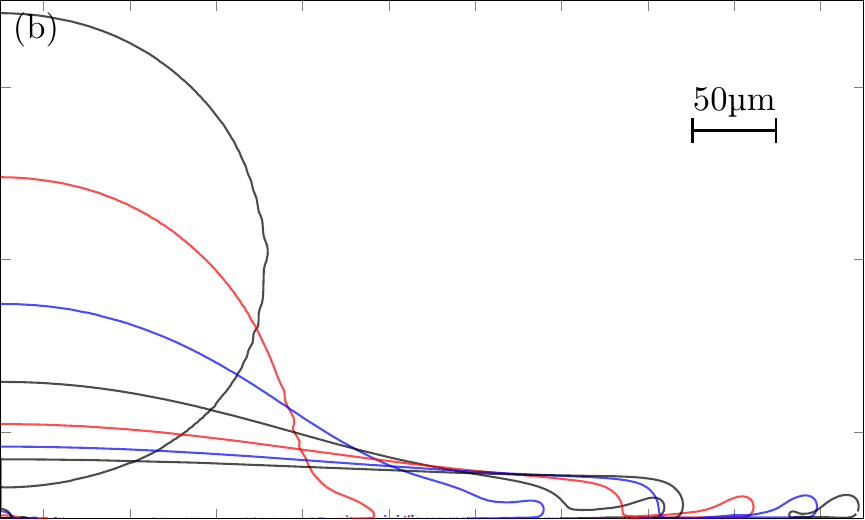}}
\caption{Time series of the interface ($\alpha = 0.5$) of a low viscosity
ethanol droplet (a) and a high viscosity silicone oil droplet (b) impacting and
spreading on a dry solid surface at atmospheric ambient pressure. Different
colors are used to be able to distinguish more clearly between successive contour
lines. The time difference between successive lines is $\Delta t/\tau = 0.0667$
for the ethanol droplet (a) and $\Delta t/\tau = 0.333$ for
the silicone oil droplet (b). Comparing both droplets shows that the
characteristic time scale for deposition, spreading, and breakup is much longer
for the high viscosity silicone oil than for the low viscosity ethanol.
}
\label{fig:contour}
\end{figure*}

\section{Theory \& Method}

\begin{figure*}[htbp]
\centering
\subfloat%
{\includegraphics[width=0.155\textwidth]{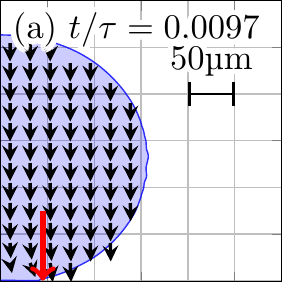}}
\hspace{4pt}%
\subfloat%
{\includegraphics[width=0.155\textwidth]{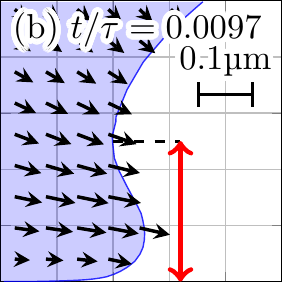}}
\hspace{4pt}%
\subfloat%
{\includegraphics[width=0.155\textwidth]{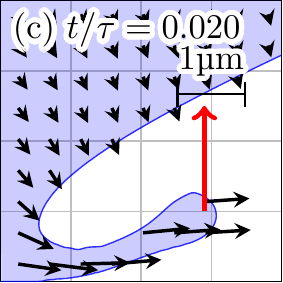}}
\hspace{4pt}%
\subfloat%
{\includegraphics[width=0.155\textwidth]{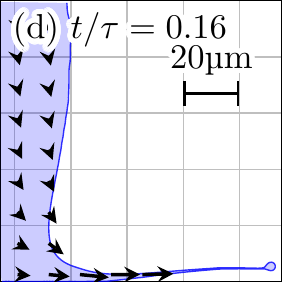}}
\hspace{4pt}%
\subfloat%
{\includegraphics[width=0.155\textwidth]{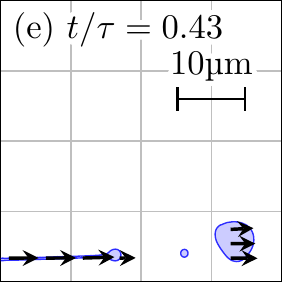}}
\hspace{4pt}%
\subfloat%
{\includegraphics[width=0.155\textwidth]{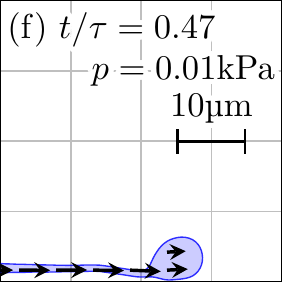}} \\
\subfloat%
{\includegraphics[width=0.155\textwidth]{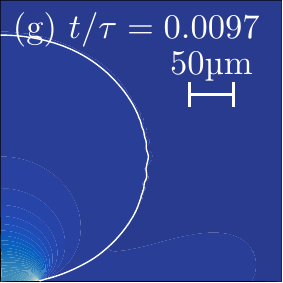}}
\hspace{4pt}%
\subfloat%
{\includegraphics[width=0.155\textwidth]{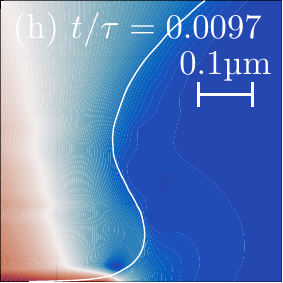}}
\hspace{4pt}%
\subfloat%
{\includegraphics[width=0.155\textwidth]{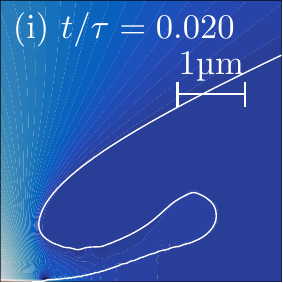}}
\hspace{4pt}%
\subfloat%
{\includegraphics[width=0.155\textwidth]{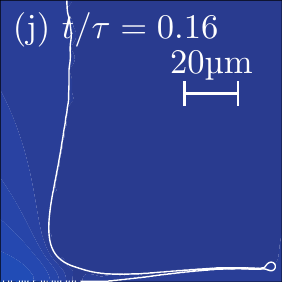}}
\hspace{4pt}%
\subfloat%
{\includegraphics[width=0.155\textwidth]{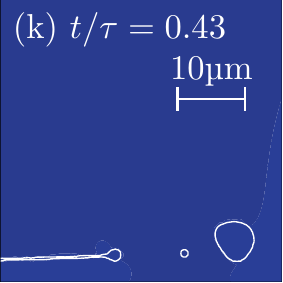}}
\hspace{4pt}%
\subfloat%
{\includegraphics[width=0.155\textwidth]{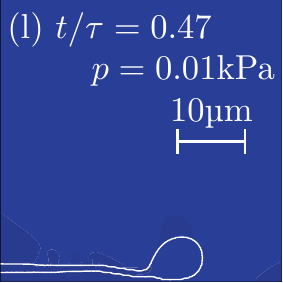}} \\
\noindent\rule{\textwidth}{1pt} \\
\subfloat%
{\includegraphics[width=0.155\textwidth]{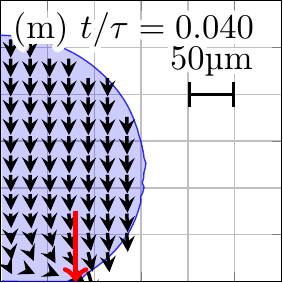}}
\hspace{4pt}%
\subfloat%
{\includegraphics[width=0.155\textwidth]{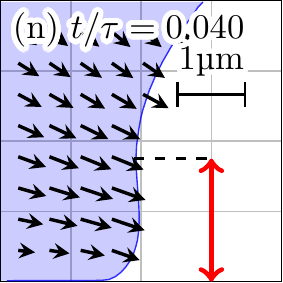}}
\hspace{4pt}%
\subfloat%
{\includegraphics[width=0.155\textwidth]{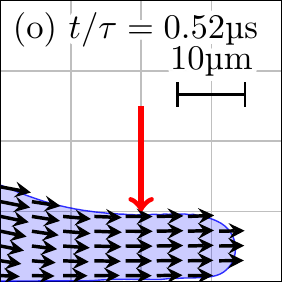}}
\hspace{4pt}%
\subfloat%
{\includegraphics[width=0.155\textwidth]{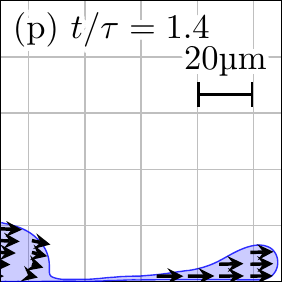}}
\hspace{4pt}%
\subfloat%
{\includegraphics[width=0.155\textwidth]{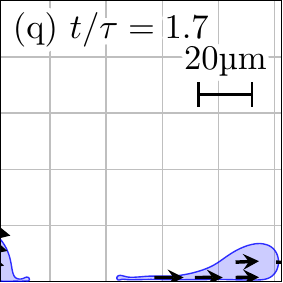}}
\hspace{4pt}%
\subfloat%
{\includegraphics[width=0.155\textwidth]{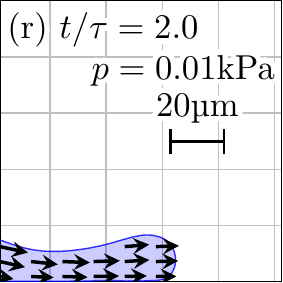}} \\
\subfloat%
{\includegraphics[width=0.155\textwidth]{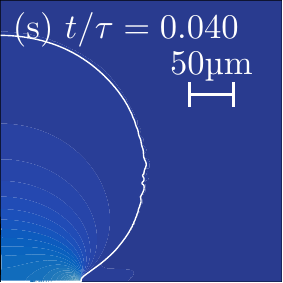}}
\hspace{4pt}%
\subfloat%
{\includegraphics[width=0.155\textwidth]{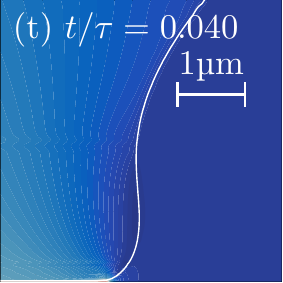}}
\hspace{4pt}%
\subfloat%
{\includegraphics[width=0.155\textwidth]{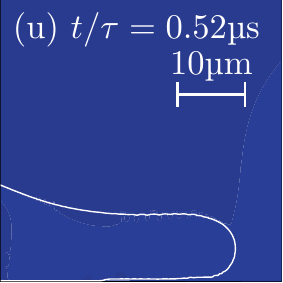}}
\hspace{4pt}%
\subfloat%
{\includegraphics[width=0.155\textwidth]{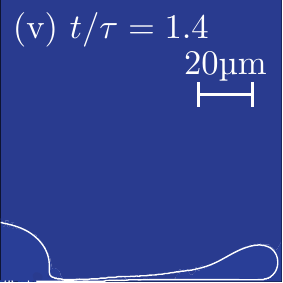}}
\hspace{4pt}%
\subfloat%
{\includegraphics[width=0.155\textwidth]{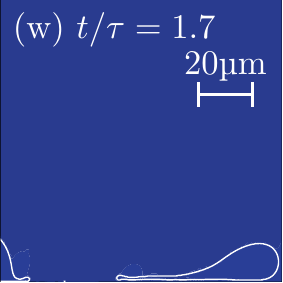}}
\hspace{4pt}%
\subfloat%
{\includegraphics[width=0.155\textwidth]{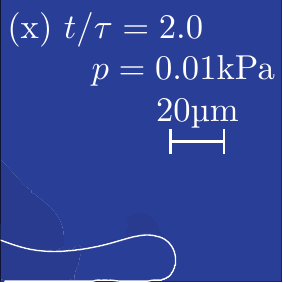}} \\
\subfloat%
{\includegraphics[width=0.49\textwidth]{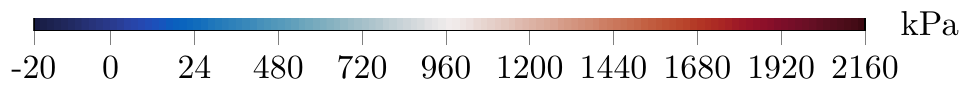}}
\caption{Time series of the impact of low viscosity ethanol droplets (a--l) and
high viscosity silicone oil droplets (m--x). In (a--f) and (m--r) blue
depicts the liquid phase and white the gas phase. The black arrows depict the
direction of the velocity vector field of the liquid phase. Images (g--l) and
(s--x) show the pressure field (in $\si{\kPa}$). All the images of the pressure
field are are shown at the same times as the vector plots.
 (a and g) Cross section of the whole ethanol droplet
at the moment of lamella formation. The red arrow in image (a) indicates the
position of the contact line.
(b and h) Magnified depiction of the contact line at the moment of lamella formation.
The height of the lamella is on the order of $0.1 \si{\mu\m}$. (c and i) Because the
lamella is very thin and travels at high speed it gets vertically ejected right
after impact as a liquid sheet. (d and j) The liquid sheet continues to travel through
the air till it breaks up at the rim in images (e) and (k). (f and l) At reduced
ambient gas pressure the liquid sheet does not get broken up and splashing is
suppressed. (m and s) Cross section of the silicone oil droplet at the moment of lamella formation. The red
arrow indicates the position of contact line. (n and t) Magnified depiction of the
contact line at the moment of lamella formation. The height of the lamella is on
the order of $1.0 \si{\mu\m}$. (o and u) Because the lamella is very thick it
continues spreading horizontally till it eventually starts thinning out at the
position of the red arrow and a liquid sheet is formed in images (p) and (v). (q and
w) At atmospheric pressure the liquid sheet breaks up while at reduced pressure the
splash is suppressed in images (r) and (x). The time series show that both the effect of ambient
gas pressure and liquid viscosity is captured in the simulations.
The observed high pressures regions are associated with gas escaping
from under the droplet upon impact and gas film collapse under the droplet.
There are no high pressure peaks associated with the spreading and breakup of
the liquid sheet. Right after breakup in images (e) and (k) the liquid sheet
thickness is about 6 grid cells.}
\label{fig:timeSeries} 
\end{figure*}
 
To be able to capture both the effect of the liquid viscosity and the ambient gas
pressure on impact, spreading, and breakup of the different droplets, the
simulations use a finite volume implementation of the Volume Of Fluid (VOF)
method \cite{hirt1981}. The VOF approach evolves around the
definition of a phase parameter $\alpha$ with the following properties:
\begin{equation}
  \alpha
=
  \left\{
  \begin{array}{ll}
  0      & \txt{in gas phase}    \\
  (0, 1) & \txt{on interface}    \\
  1      & \txt{in liquid phase}
  \end{array}
  \right.
\end{equation}
The evolution of $\alpha$ is calculated using the following transport equation:
\begin{equation}
  \frac{\partial \alpha}{\partial t}
+ \nabla \cdot \brc{\alpha \vec{v}}
+ \nabla \cdot \brc{\alpha \brc{1 - \alpha} \vec{v}_{lg}}
=
  0,
\end{equation}
where $\vec{v}$ is the phase averaged velocity, and $\vec{v}_{lg}$ is a velocity
field suitable to compress the interface. This equation is equivalent to a
material derivative, but rewritten to minimize numerical diffusion
\cite{rusche2002}.

The phase parameter is used to calculate the phase averaged density, $\rho$,
velocity, $\vec{v}$, and viscosity, $\mu$, which are used in the momentum balance:
\begin{equation}
  \frac{\partial \rho \vec{v}}{\partial t}
+ \nabla \cdot \brc{\rho \vec{v} \otimes \vec{v}}
=
- \nabla p
+ \nabla \cdot \brc{\mu \nabla \vec{v}}
+ \rho \vec{g}
- \vec{f},
\end{equation}
and the continuity equation:
\begin{equation}
  \nabla \cdot \vec{v} 
= 
  0.
\end{equation}
In the above equations $t$ is time, $p$ is pressure, $g$ is gravity, $\vec{f}$ is any body force, like
the surface tension force, and $\otimes$ is the dyadic product. To complete the
VOF model, an expression is needed to calculate the surface tension force
$\vec{f}_{\txt{st}}$, and a model is needed for the contact line. The
surface tension force is calculated using the expression \cite{brackbill1992}:
\begin{equation}
  \vec{f}_{\txt{st}}
=
  \sigma_{\txt{st}} \kappa \nabla{\alpha}
\end{equation}
where $\sigma_{\txt{st}}$ is the surface tension coefficient, and $\kappa$ is
the curvature of the interface. 

The effect of varying the Young's angle $\theta_{0}$ from $\ang{0}$ to
$\ang{180}$ is calculated directly through the Generalized Navier Boundary
Condition (GNBC) at the impact wall \cite{qian2003,gerbeau2009}. With this boundary
condition the dynamic contact angle $\theta$ is allowed to vary freely, but a
restoring line-tension force is applied at the contact line whenever the dynamic
angle deviates from $\theta_{0}$. This restoring force is an additional source
term in the Navier-Stokes equations, and has the following form:
\begin{equation}
  \vec{f}_{\txt{lt}}
=
- \frac{\sigma_{\txt{st}}}{h}
  \cos{\theta_{0}}
  \nabla_{\txt{2D}}{\alpha}
\end{equation}
In the above equation $\sigma_{\txt{st}}$ is the surface tension coefficient,
$h$ is the height of the local grid cell, and $\nabla_{\txt{2D}}{\alpha}$ is the
gradient of $\alpha$ on the wall. This force is applied on the liquid-gas
interface in the first grid cells adjacent to the wall and is balanced by the
surface tension force when $\theta$ is equal to $\theta_{0}$. Away from the
contact line the used implementation of the generalized Navier boundary
condition reduces to the Navier-slip boundary condition. Using this slip 
boundary condition gives a good approximation for the thin film behavior at the
wall \cite{lauga2007,sprittles2017}. Because the model used can accommodate only
one value for the slip length, a value of $\lambda = 1 \si{\nano\m}$ is chosen to
be able to accurately describe the contact line. However, in practice the
effective slip length is on the order of the mesh size of $10 \si{\nano\m}$
\cite{jacqmin2000}. This results in the gas film potentially closing faster in
our simulations than if the slip length were truly $1 \si{\nano\m}$. 

\renewcommand{\arraystretch}{1.1}
\begin{table}[h]
\caption{An overview of the non-dimensional numbers and ratios of liquid and gas
properties at atmospheric pressure. For the simulations at reduced pressure the
dynamic viscosity is kept constant, but the kinematic
viscosity and gas density are allowed to change.}
\label{tab:nondim}
\begin{tabular}{lrrrr}
                  & $\txt{Re}$ & $\txt{We}$ & $\rho_{l}/\rho_{g}$ & $\nu_{l}/\nu_{g}$ \\
\hline
\hline
Ethanol           & $1973$     & $1057$     & $789$               & $0.1028$          \\
\hline                                                                          
Silicone oil      & $281$      & $1396$     & $935$               & $0.7226$          \\
\hline
\hline
\end{tabular}
\end{table}

The simulations are performed for ethanol and silicone oil in air using the VOF
solver of the OpenFOAM Finite Volume toolbox \cite{openfoam} at up to $10
\si{\nano \m}$ resolution at the wall. Complete convergence at the contact line
would require a grid size below the slip length,
which is beyond the reach of our computational resources. Nevertheless, at a
grid size of $10 \si{\nano\m}$ the necessary physics of splashing are already
present, and we expect the main observations of our simulations to be
qualitatively correct \cite{boelens2016a}. To reduce memory requirements, the
simulations are performed in a 2-D axisymmetric geometry and the
droplets have a diameter of $300 \si{\micro \m}$. This results in the
non-dimensional numbers shown in Tab.~\ref{tab:nondim}. The Reynolds number is
defined as: $\txt{Re} = V_{0} D/\nu_{l}$, and the Weber number as: $\txt{We} =
\rho_{l} V_{0}^{2} D/\sigma$. The material properties of ethanol and silicone
oil were chosen because these liquids have been used in many experiments
\cite{xu2005,xu2007b,driscoll2010,stevens2014b} and because the defining
difference between them is their viscosity. For the simulations at reduced ambient gas
pressure the density of the gas phase is reduced $100$ times while keeping the
dynamic viscosity constant \cite{kadoya1985}. This value is well below the
ambient pressure threshold typically observed in experiments \cite{xu2005} to
make sure the simulations are performed well into the suppressed splashing regime.
More information on the boundary conditions at the contact line can be found in
Ref.~\citenum{boelens2016b}. More information on the equations, initial conditions, and a comparison with
experiments van be found in Ref.~\citenum{boelens2016a}. A direct comparison
between simulations and experiments is very challenging because the simulations
cannot be scaled up due to computational costs and the experiments are hard to
scale down due to the difficulties of imaging micro meter scale droplets
\cite{visser2015}. However, in this paper an indirect comparison is performed
and it is shown that the scaling of the gas film height as function of impact velocity is
consistent with theory and experiments. Also, multiple experimental observations
are reproduced, including the formation of the central air bubble, liquid sheet
formation, and contact line instability \cite{boelens2016a}.

\section{Results}

Unless mentioned otherwise, all units are made dimensionless using the impact
velocity $V_{0}$ and droplet diameter $D$. This results in an inertial time
scale: $\tau = D/V_{0} = 30 \si{\micro\s}$. As a droplet spreads over a surface, two different gas
films are observed under the droplet; initially at the edge of the droplet a
very thin gas film is present, on the order of $10 \si{\nano \m}$ thick
\cite{mandre2009,kolinski2012,boelens2016a}. When the liquid spreads on top of
this gas film in a rolling motion \cite{philippi2016,latka2018} or when there is
a 3 phase contact line present this is called a lamella. When the liquid spreads
on a gas film which is on the order of several $\si{\micro \m}$ thick
\cite{driscoll2010,boelens2016a} or when the liquid gets lifted up in the air
completely this is called a liquid sheet. At the resolution used in this work
the gas film on the order of $10 \si{\nano \m}$ is under-resolved. However, in
previous work it was shown that as long as the gas film is present all the
essential features of splashing are captured \cite{boelens2016a}.

To illustrate the difference between a low and high viscosity splash,
Fig.~\ref{fig:contour} shows a time series of the interface of both a simulated
ethanol droplet (a) and a simulated silicone oil droplet (b) splashing at
atmospheric ambient pressure. Different colors are used to improve readability
of the plots. In Fig.~\ref{fig:contour}~(a) the time difference between two curves is $\Delta
t/\tau = 0.0667$ and in Fig.~\ref{fig:contour}~(b) the time difference is $\Delta t/\tau = 0.333$.
A first observation that can be made by comparing both time series
is how much longer the silicone oil droplet takes to splash than the ethanol
droplet. Also, the lamella and subsequent liquid sheet formed in the ethanol
splash appears to be much thinner than the lamella formed in the silicone oil
splash, which is consistent with literature \cite{mongruel2009}. On the other hand,
the typical crown splash observed in experiments when an ethanol droplet impacts
on a solid dry wall is not observed in the simulations. This could be caused by
the fact that the droplets used in these simulations are much smaller than a
typical droplet used in experiments resulting in a smaller liquid sheet and thus
smaller lift force acting on the liquid \cite{boelens2016a}. To further investigate this difference
between the experimental and simulation results, a closer look is taken at the
simulations in Fig.~\ref{fig:timeSeries}.

Fig.~\ref{fig:timeSeries} shows a time series of the impact of low viscosity
ethanol droplets (a--l) and high viscosity silicone oil droplets (m--x). In the
images (a--f) and (m--r) the liquid phase is blue and the gas phase is white. The black
arrows show the direction of the velocity vector field inside the liquid. In
addition to the direction of the velocity vector field,
Figs.~\ref{fig:timeSeries}~(g--l)~and~(s--x) show the pressure field at the same
times and locations as the vector field. Figs.~\ref{fig:timeSeries}~(a)~and~(g) show the whole ethanol
droplet at the moment a lamella can first be detected at atmospheric ambient
pressure. Subsequent images are zoomed in at the contact line, where the red
arrow is pointing in Fig.~\ref{fig:timeSeries}~(a).
Figs.~\ref{fig:timeSeries}~(m)~and~(s) also shows the whole droplet at the
moment of lamella ejection, but for silicone oil. Comparing
Figs.~\ref{fig:timeSeries}~(a)~and~(g) with
Figs.~\ref{fig:timeSeries}~(m)~and~(s) shows again that the dynamics
are much slower for high viscosity liquids and that for silicone oil the lamella
forms much later than for ethanol. In Figs.~\ref{fig:timeSeries}~(b)~and~(h) and
Figs.~\ref{fig:timeSeries}~(n)~and~(t) both droplets are shown at the same times as in
Figs.~(a)~and~(g) and Fig.~(m)~and~(s), but zoomed in at the contact line. The shape of the
droplets looks very similar, but the characteristic length scale of the lamella
is much larger for silicone oil than ethanol. This is consistent with literature
that proposes self-similarity solutions to describe the interface of the spreading
droplets \cite{philippi2016}. As the droplets approach the surface a large pressure build up can be observed
at the stagnation point. This causes the droplet to deform and a new high
pressure region forms away from the center where the ambient gas is escaping
from under the droplet at high speed. This high pressure area is
still present at the time of lamella ejection in
Figs.~\ref{fig:timeSeries}~(g)~and~(h). In Fig.~\ref{fig:timeSeries}~(t) also a higher pressure can be
observed at the interface under the droplet. This is caused by the break down of
the gas film under the droplet into small gas bubbles. This causes high local
curvature of the interface and corresponding regions of high Laplace pressure.
In the case of the ethanol droplet a thicker gas
film forms right away at the edge of the lamella, turning it into a liquid sheet.
In addition, this liquid sheet gets ejected with a strong vertical component,
similarly to the behavior observed in a crown splash. For the silicone oil droplet
the lamella continues to spread horizontally till eventually the lamella starts
thinning out. This is shown in Figs.~\ref{fig:timeSeries}~(o)~and~(u) where the red arrow
indicates the location where a minimum in the interface is first observed.
Figs.~\ref{fig:timeSeries}~(d)~and~(j) show the ethanol liquid sheet continuing to travel through the air
till eventually the sheet breaks up at the edge of the rim in
Figs.~\ref{fig:timeSeries}~(e)~and~(k). The silicone oil lamella meanwhile continues to thin out and a thicker air film is entrained under
the lamella, turning it into a liquid sheet in
Figs.~\ref{fig:timeSeries}~(p)~and~(v). In Figs.~\ref{fig:timeSeries}~(q)~and~(w) the
silicone oil liquid sheet breaks up. However, in this case the sheet breaks up at
the lamella side and not at the rim, suggesting a different breakup mechanism.
Figs.~\ref{fig:timeSeries}~(f)~and~(l) and Figs.~\ref{fig:timeSeries}~(r)~and(x) show the ethanol and
silicone oil liquid sheet staying intact and not breaking up in the case of
simulations with a reduced ambient gas pressure.
 As can be seen in Figs.~\ref{fig:timeSeries}~(j--l) and
Figs.~\ref{fig:timeSeries}~(u--x), there are no high pressure regions observed
associated with the spreading of the lamella and liquid sheet breakup.

\begin{figure}[htbp]
\centering
\includegraphics[width=0.49\textwidth]{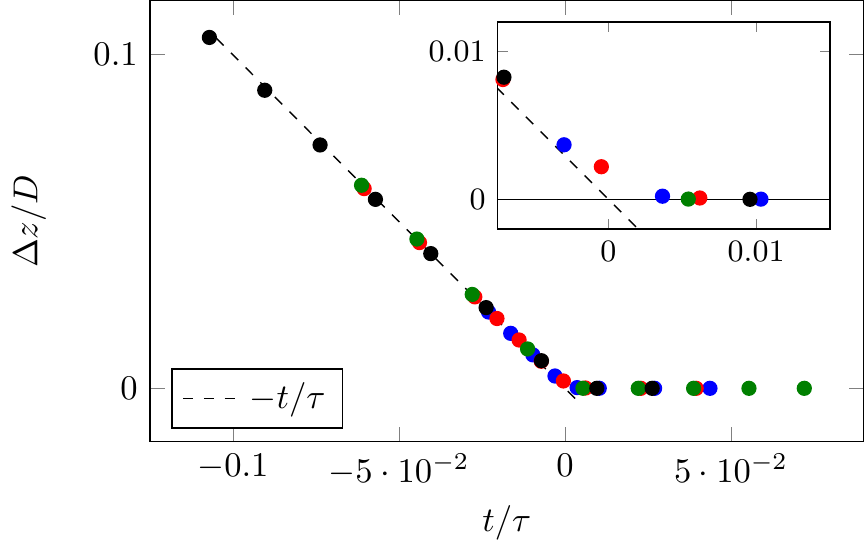}
\caption{The height between the bottom of the droplet and the surface, $\Delta
z$, as function of time, $t$, for 
silicone oil at reduced ambient pressure
(\protect\includegraphics{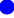}),
silicone oil at normal ambient pressure
(\protect\includegraphics{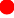}),
ethanol at reduced ambient pressure
(\protect\includegraphics{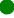}), and
ethanol at normal ambient pressure
(\protect\includegraphics{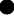}). Impact time, $t = 0$, is defined as the moment the droplet would have
hit the surface if no gas film formed under the droplet. This trajectory is
described by the line $\Delta z/D = - t/\tau$. The inset shows the deviation
from the impact trajectory due to gas film formation.}
\label{fig:impactTime}
\end{figure}
When a droplet impacts a solid wall initially a thin gas film forms under the
droplet, preventing the liquid from
touching down on the surface \cite{mandre2009,kolinski2012,boelens2016a}.
Therefore the moment that liquid can first be detected
on the wall is not a good definition for the moment of impact. Instead, the
height between the bottom of the droplet and the wall, $\Delta z$, is
plotted as a function of time, $t$, and a straight line is fitted to the
trajectory of the droplet right before impact. As can be seen in
Fig.~\ref{fig:impactTime}, impact time, $t = 0$, is then defined as the
moment the droplet would have hit the wall if the gas film under the droplet did
not form. The inset shows the deviation from the impact trajectory due to gas
film formation.

\begin{figure*}[htbp]
\centering
\subfloat%
{\includegraphics[width=0.49\textwidth]{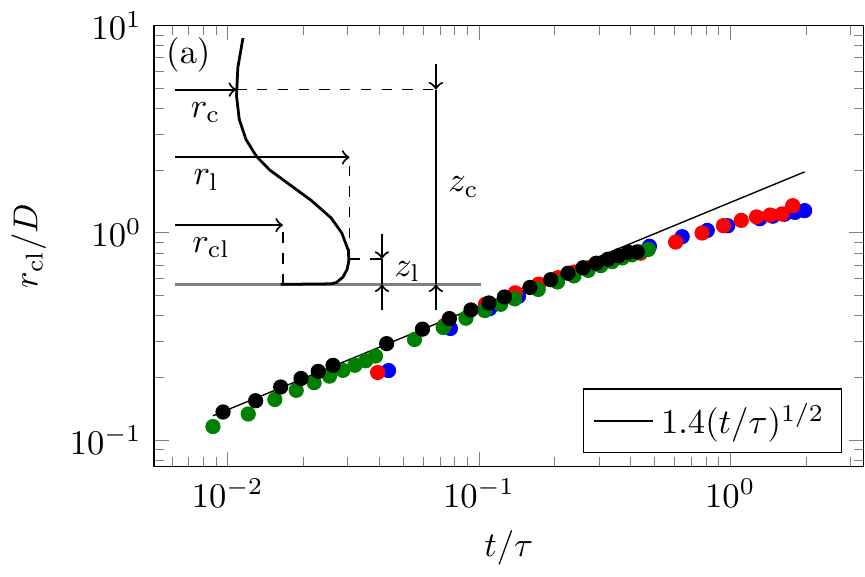}}
\hspace{5pt}%
\subfloat%
{\includegraphics[width=0.49\textwidth]{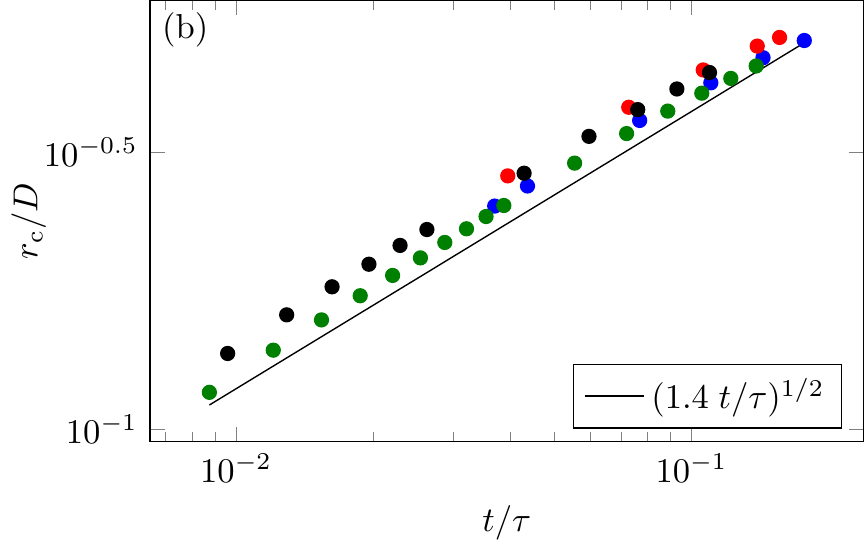}} \\
\subfloat%
{\includegraphics[width=0.49\textwidth]{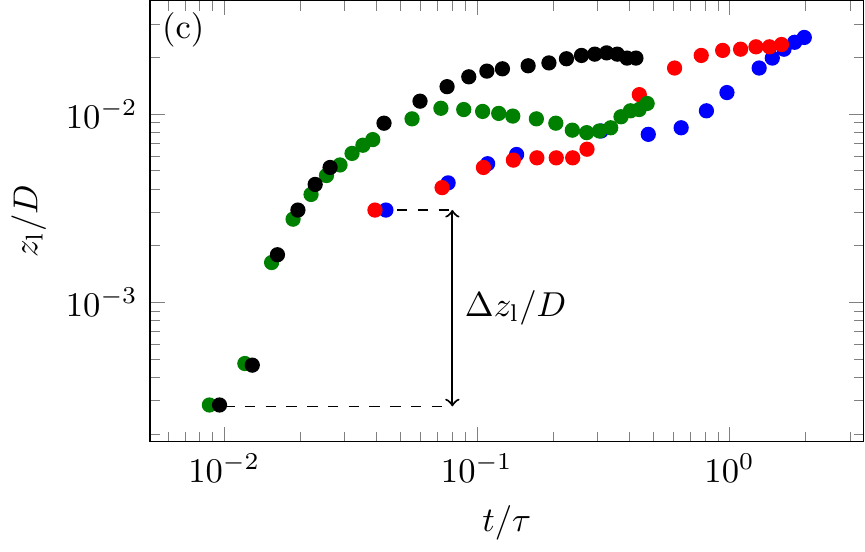}}
\hspace{5pt}%
\subfloat%
{\includegraphics[width=0.49\textwidth]{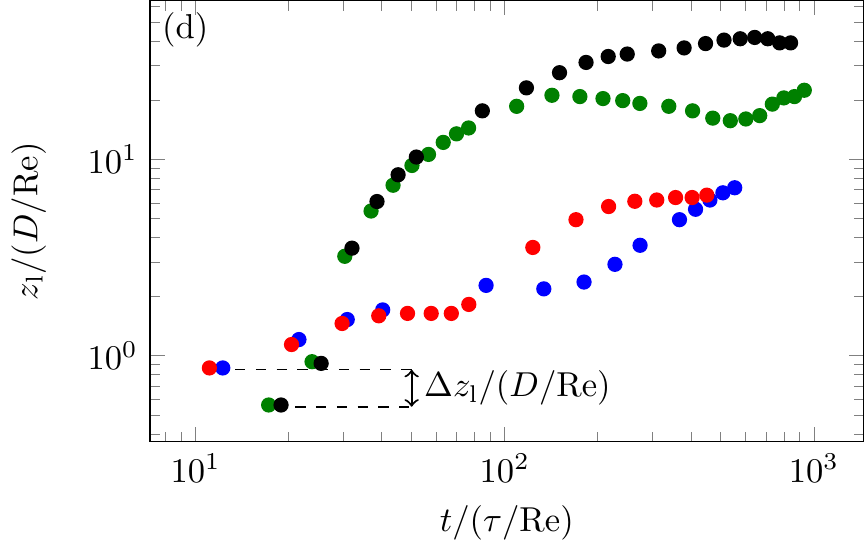}} \\
\subfloat%
{\includegraphics[width=0.49\textwidth]{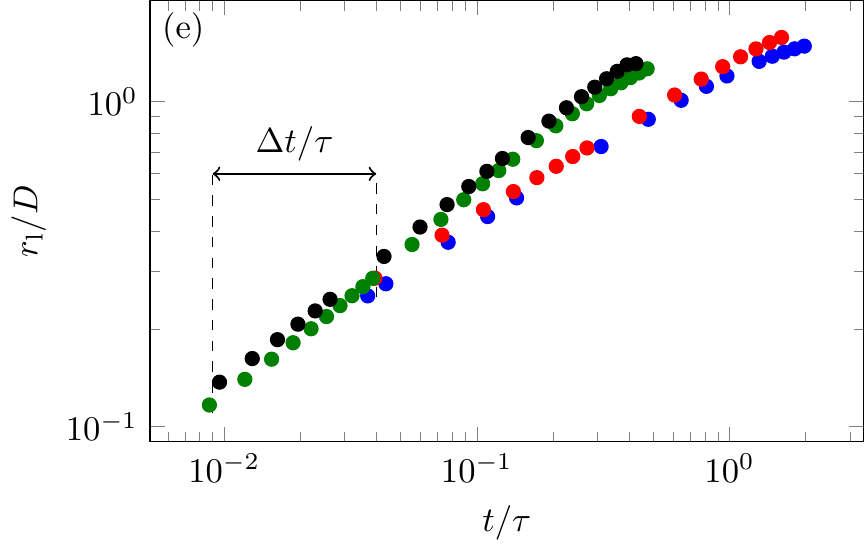}}
\hspace{5pt}%
\subfloat%
{\includegraphics[width=0.49\textwidth]{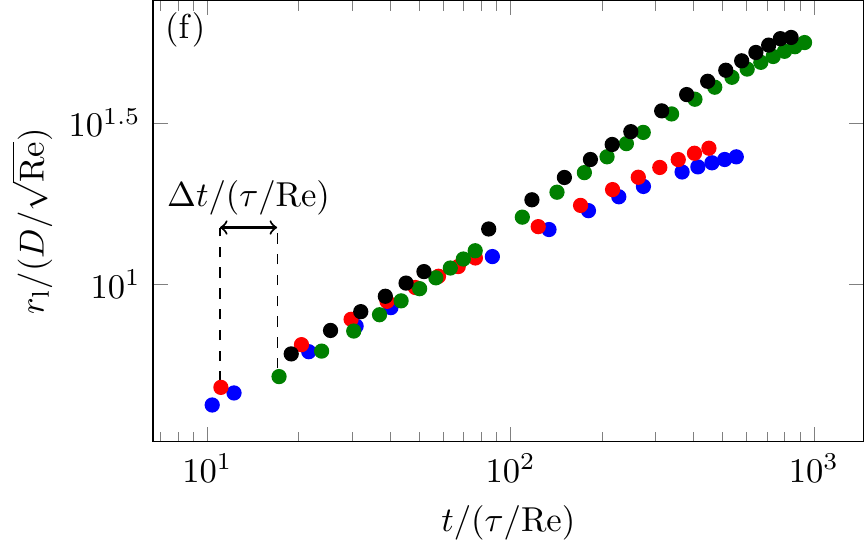}}
\caption{%
The insert in frame (a) shows a schematic drawing of the parameters referred to in the main text.
$r_{\txt{c}}$, and $z_{\txt{c}}$ are the radial position and height,
respectively, of the cusp which forms as the droplet spreads over the surface.
$r_{\txt{cl}}$ is the radial position of the contact line, and $r_{\txt{l}}$,
and $z_{\txt{l}}$ show the radial position and height of the lamella, respectively.
(a) The radial position of the contact line, $r_{\txt{cl}}$, as function of time for  
(\protect\includegraphics{sil001})
silicone oil at reduced ambient pressure,
(\protect\includegraphics{sil100})
silicone oil at atmospheric ambient pressure,
(\protect\includegraphics{eth001}) 
ethanol at reduced ambient pressure, and
(\protect\includegraphics{eth100}) 
ethanol at atmospheric ambient pressure.
(b) The radial position of the cusp, $r_{\txt{c}}$, as function of time.
(c) The height of the lamella/liquid sheet, $z_{\txt{l}}$, as function of time
using inertial scaling.
(d) The height of the lamella/liquid sheet, $z_{\txt{l}}$, as function of time
using viscous scaling.
(e) The radial position of the lamella, $r_{\txt{l}}$, as function of time using
inertial scaling.
(f) The radial position of the lamella, $r_{\txt{l}}$, as function of time using
mixed scaling.
The theoretical curves in both image (a) and (b) are proposed by \citet{mongruel2009}, and show that both length scales scale with inertia.
When using inertial scaling in image (c) no collapse of the data can be observed.
However, When viscous scaling is applied in image (d) at early times the lamella
thickness for ethanol and silicone oil becomes of the same order.
When applying inertial scaling in image (e) a large time delay can
be observed between the moment that the lamella can first be detected in the
case of the low viscosity ethanol and the high viscosity silicone oil. When
applying the scaling suggested by \citet{mongruel2009} in image (f) this time delay is
greatly reduced. The error bars in this figure are smaller than the symbols.}
\label{fig:mongruel}
\end{figure*}

After impact the droplet spreads over the surface and, when the impact velocity
is high enough, eventually a lamella forms. According to \citet{mongruel2009}
for low surface tension liquids the moment of lamella formation and the thickness
of the lamella are determined by a the balance between inertial and viscous
forces. The inset in Fig.~\ref{fig:mongruel}~(a) shows some of the different length scales identified by
\citet{mongruel2009} which are also used in this
work: $r_{\txt{c}}$, and $z_{\txt{c}}$ are the radial position and height
, respectively, of the cusp which forms as the droplet spreads over the surface.
This cusp is defined a the minimum of the interface in the radial direction.
$r_{\txt{l}}$, and $z_{\txt{l}}$ show the radial position and height of the
lamella. This position is defined by the maximum of the interface in the radial
direction. At the moment of lamella formation $r_{\txt{c}} \approx r_{\txt{l}}$
and $z_{\txt{c}} \approx z_{\txt{l}}$, and at early times the height $z_{\txt{l}}$ is the characteristic length
scale of the lamella. However, in the case of ethanol, at later times when the
liquid sheet is ejected into the air, $z_{\txt{l}}$ is a measure of how high
the liquid sheet travels above the surface. $r_{\txt{cl}}$ is the radial
position of the contact line.

 The characteristic length scale for the impact regime is the diameter of
the droplet, and thus the corresponding Reynolds number is large. As a results
the impact regime is dominated by inertia and, because of geometrical
considerations, the spreading radius of a droplet scales as: $r_{\txt{cl}} / D
\propto \sqrt{t/\tau}$ \cite{rioboo2002}. Fig.~\ref{fig:mongruel}~(a)
shows the spreading radius of the ethanol and silicone oil droplets at both
atmospheric and reduced ambient pressure as function of time. The observed
scaling is consistent with literature \cite{rioboo2002,mongruel2009}. 
The radial position of the cusp is shown in
Fig.~\ref{fig:mongruel}~(b). Because the location of the cusp closely follows a
trajectory on the original spherical interface of the impacting droplet, the
position of the cusp also scales with inertia. The relation $r_{\txt{c}}/D = 1.4
(t/\tau)^{1/2}$ is proposed by \citet{mongruel2009} and is consistent with the
simulation results.

On the other hand, the height of a lamella, $z_{\txt{l}}$, is a very small length
scale resulting in a small Reynolds number and flow being dominated by viscous
forces. Fig.~\ref{fig:mongruel}~(c) shows the evolution of the lamella height,
$z_{\txt{l}}$, as function of time, $t$, for both ethanol and silicone oil. In the case of an
ethanol droplet, right after the lamella is ejected it is lifted up in the air
and $z_{l}$ increases accordingly. However, when the appropriate scaling is
chosen, at the moment of lamella formation the data for both liquids should
collapse. Considering the difference between the lamella height for ethanol and
silicone oil, $\Delta z_{\txt{l}}$, it can be observed
that the lamella height does indeed not scale with inertia. On the other hand,
when viscous scaling is applied in Fig.~\ref{fig:mongruel}~(d) the data
collapses for the moment of lamella ejection and the height of the lamella for
both liquids is of the same order. The balance between inertial and viscous
forces determines the time of lamella ejection and Fig.~\ref{fig:mongruel}~(e)
shows that increasing the viscosity of the liquid delays the time of lamella
ejection, $\Delta t$, significantly. By applying the scaling proposed by
\citet{mongruel2009} which takes into account both inertial and viscous forces
the data can be made to collapse. In Fig.~\ref{fig:mongruel}~(f) the time of
lamella ejection is on the same order for both ethanol and silicone oil. Apart
from scaling considerations, another feature that can be be observed in both
Fig.~\ref{fig:mongruel} (c) and Fig.~\ref{fig:mongruel} (d) is that the ambient
gas pressure affects the height of the lamella, $z_{l}$. In the case of ethanol
this is a difference in the height at which the lamella travels above the
surface, which has been proposed as an explanation for the pressure effect
\cite{riboux2014,boelens2016a}. For the silicone oil simulations the height
difference is mostly caused by a difference in the shape of the lamella/liquid
sheet. This difference in shape is now further explored.

\begin{figure}[htbp]
\centering
{\includegraphics[width=0.49\textwidth]{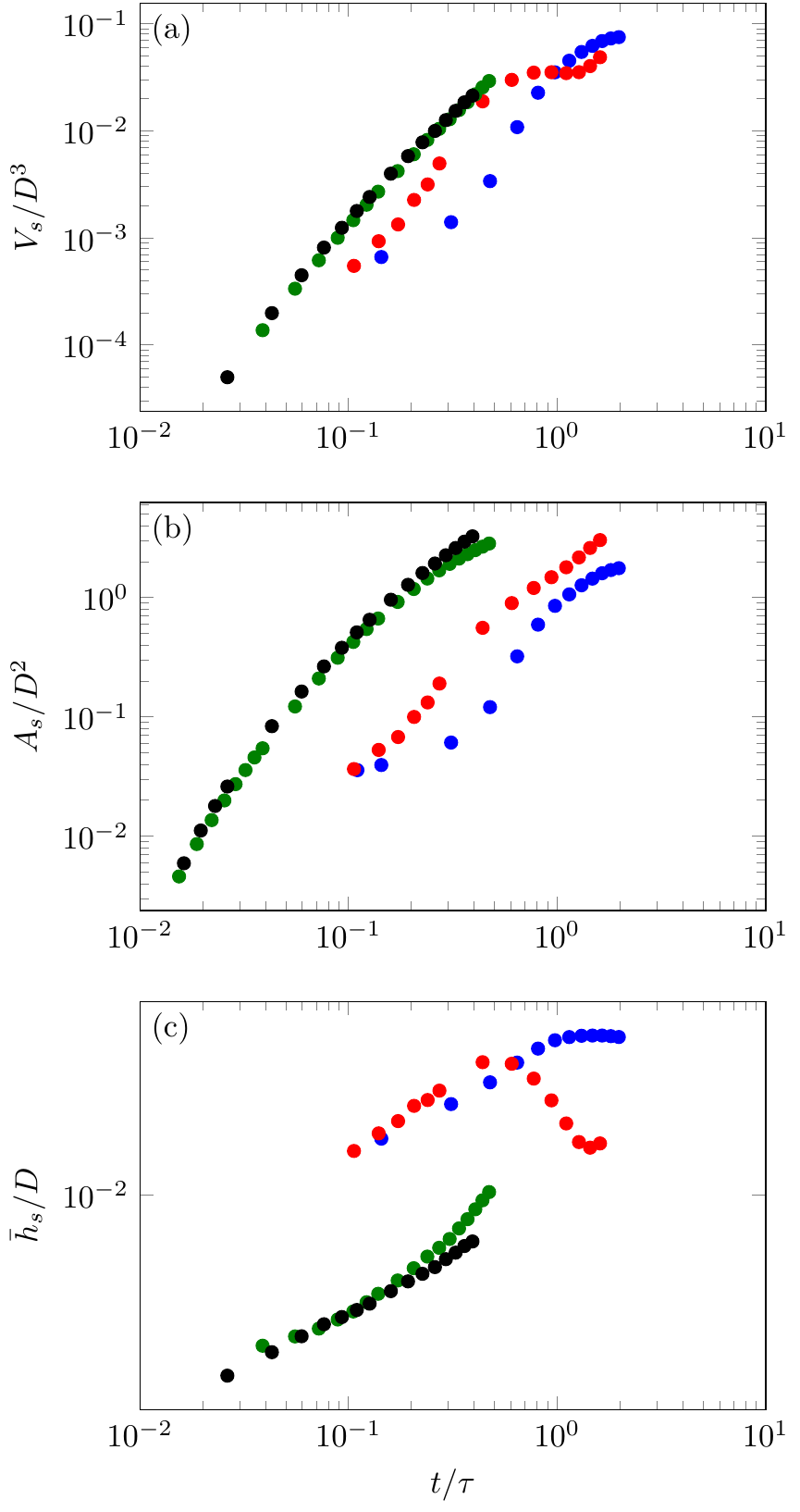}}
\caption{(a) The volume occupied by the liquid in the liquid sheet for
(\protect\includegraphics{sil001})
silicone oil at low ambient pressure, 
(\protect\includegraphics{sil100})
silicone oil at normal ambient pressure,
(\protect\includegraphics{eth001})
ethanol at low ambient pressure, and
(\protect\includegraphics{eth100})
ethanol at normal ambient pressure. 
The influx of liquid for the silicone oil
liquid sheet at atmospheric ambient pressure is initially larger than at reduced
ambient pressure,. However, later in time a significant reduction can be observed. (b) The
surface area covered on the wall by the liquid sheet. Because the contact line
moves faster at reduced ambient pressure the surface area covered by the liquid
sheet is smaller for both ethanol and silicone oil. (c) The average height of
the liquid sheet drops significantly for silicone oil at atmospheric ambient
pressure. This causes the liquid sheet to break
up. These dynamics are not present for ethanol.}
\label{fig:lamella}
\end{figure}

In addition to decreasing the gas film thickness under the liquid sheet, lowering of the ambient gas pressure also
affects the position of the contact line, $r_{\txt{cl}}$, and
the position of the edge of the droplet/lamella, $r_{\txt{l}}$. These two effects combined results in a
smaller surface area, $A_{s} = \pi \brc{r_{\txt{l}}^{2} - r_{\txt{cl}}^{2}}$,
being covered by the liquid sheet. While this effect is very small for the
ethanol droplets, for the simulations of silicone oil droplets this effect is
quite pronounced, as can be seen in Fig.~\ref{fig:lamella}~(b). The difference
in the area covered by the liquid sheet also causes the volume of the liquid
sheet, $V_{s}$, to be different. This volume is defined as the space occupied by the
liquid between  the contact line and the edge of the droplet/lamella: $r_{\txt{cl}} > r > r_{\txt{l}}$. For ethanol the volume
of the liquid sheet does not depend on the pressure. However,
Fig.~\ref{fig:lamella}~(a) shows that there is a clear difference for the
silicone oil simulations. Initially there is a larger liquid volume present in
the liquid sheet at atmospheric ambient pressure. At later times the inflow of
liquid is greatly reduced. This is partly because there is a positive feedback
loop for the average liquid sheet thickness, $\bar{h}_{s} = V_{s}/A_{s}$. In Fig.~\ref{fig:lamella}~(c) it can be
observed that while initially the average thickness of the silicone oil liquid sheet is
larger at atmospheric ambient pressure, eventually the influx of liquid cannot
keep up with the spreading rate of the droplet and the average height decreases.
This decrease in turn further limits the inflow of liquid, and the average
height further declines till the liquid sheet breaks up. 

\section{Discussion}

The simulations confirm the existence of a thin gas film at the edge of the
spreading droplet \cite{mandre2009,kolinski2012}. A more in-depth analysis of the
behavior of this gas film as function of different material properties is found
in Ref.~\citenum{boelens2016a}. In this paper a closer look is taken at the
properties of the gas film right before lamella formation and it is found that
the simulations are able to reproduce the scaling of the gas film found in
literature \cite{mandre2012}. While the simulations presented in this work are
performed in the continuum limit, the observed difference in the contact line
velocity between atmospheric and reduced ambient pressure is consistent
with the work of \citet{sprittles2017}. Since \citet{mandre2009} did not
investigate subcontinuum effects in the gas film, this could provide an
explanation of why they were not able to find a dependence of the gas film
height on the ambient gas pressure. As shown in Fig.~\ref{fig:mongruel}, the
scaling of the lamella ejection time and the scaling of the lamella height as
function of viscous and inertial forces is consistent with the work of
\citet{mongruel2009}. This is in agreement with earlier work where we find a
collapse of the data at the moment of lamella formation for a larger range of
impact velocities \cite{boelens2016a}.

Unfortunately, the computational cost of these simulations make it unfeasible to perform
simulations for a larger number of different viscosities: a simulation of a low
viscosity ethanol splash takes several weeks and the simulation of a
high viscosity silicone oil splash takes several months to complete. Therefore it was not
possible to investigate the origins of the different trends which are observed
when plotting the splashing threshold pressure as function of viscosity
\cite{xu2007b,driscoll2010,stevens2014b}. However, our observations concerning the thin gas film and
lamella behavior are made for both low viscosity and high viscosity liquids.
This confirms that the early time stages of droplet deposition are identical between 
high and low viscosity liquids and that thus conclusions reached from
research on high viscosity liquids should also be valid for low viscosity
liquids.

For late time spreading and breakup behavior of the droplets on the other hand
significant differences are found. While at first glance the simulations do not
seem to reproduce a crown splash, Fig~\ref{fig:timeSeries} shows that in the
case of a low viscosity liquid a very thin lamella forms right after impact and
is ejected into the air at a high velocity. This causes a thicker air film to
form under the liquid right away, and the lamella gets ejected as a liquid
sheet. This behavior of the low viscosity liquid is also observed in experiments
\cite{xu2005} and is consistent the idea of a lifting force acting on the
liquid sheet \cite{riboux2014,riboux2015,riboux2017}. As shown in
Fig.~\ref{fig:lamella} the liquid sheet for the high viscosity liquid forms in a
different manner. A thick lamella becomes a thin liquid sheet due to limited
inflow of liquid into the sheet. The identification of these two different types
of liquid sheet formation and breakup is consistent with literature
\cite{xu2007b,driscoll2010,stevens2014b} and confirms that theories on splashing
should take viscosity into consideration as an important parameter in their
models. 

\section{Conclusions}
In this work simulations are presented of low viscosity ethanol and high
viscosity silicone oil droplets impacting on a dry solid surface at atmospheric
and reduced ambient pressure. To account for the liquid and gas phase the Volume
Of Fluid (VOF) approach is used together with the Brackbill surface tension model \cite{brackbill1992} and the
Generalized Navier Boundary Condition \cite{qian2003,gerbeau2009} to describe the surface tension and
contact line behavior, respectively. The simulations are able to capture both
the effect of the ambient gas pressure and liquid viscosity on droplet impact
and breakup. The results suggests that the early time impact and gas film
behavior for both low and high viscosity liquids share the same physics.
However, for later time liquid sheet formation and breakup low and high
viscosity liquids behave differently. A low viscosity lamella gets ejected into
the air right after formation and becomes a liquid sheet. This is consistent
with the idea of a lift force acing on the liquid \cite{riboux2014,riboux2015,riboux2017}. A high viscosity liquid sheet
on the other hand forms due to a limited inflow of liquid into the lamella/liquid sheet later in the spreading stage of the droplet. These results explain
why for both kind of liquids the pressure effect can be observed, while at the
same time different splashing regimes can be
identified\cite{xu2007b,driscoll2010,stevens2014b}. For future work we propose
to closer investigate the transition between the low and high viscosity
splashing regime to better understand the behavior of the threshold pressure as
function of viscosity.

\begin{acknowledgments}
The authors would like to thank Sid Nagel and Andrzej Latka for many fruitful
discussions and great insights.
\end{acknowledgments}

%

\end{document}